\documentclass[lettersize,journal]{IEEEtran}
\usepackage{amsmath,amsfonts}
\usepackage{algorithmic}
\usepackage{algorithm}
\usepackage{array}
\usepackage[caption=false,font=normalsize,labelfont=sf,textfont=sf]{subfig}
\usepackage{textcomp}
\usepackage{stfloats}
\usepackage{url}
\usepackage{verbatim}
\usepackage{graphicx}
\usepackage{cite}
\begin{document}

\title{Proportional-Fair Joint User Grouping and Power
Allocation for Uplink NOMA-ISAC}

\author{Yaxuan Luo
\thanks{Y. Luo is graduated from the School of Engineering in 2025, The University of Manchester, Manchester, United Kingdom (email: nancie7luoyaxuan@gmail.com).}}

\markboth{Journal of \LaTeX\ Class Files,~Vol.~14, No.~8, August~2021}%
{Shell \MakeLowercase{\textit{et al.}}: A Sample Article Using IEEEtran.cls for IEEE Journals}

\IEEEpubid{}

\maketitle

\begin{abstract}
This letter addresses long-term fairness in uplink non-orthogonal multiple access integrated sensing and communication (NOMA-ISAC) systems. Existing resource allocation schemes that maximize instantaneous sum rate often favor strong users, leaving historically underserved users with poor long-term throughput. We propose PF-JUGPA, a proportional-fair scheduling based joint user grouping and power allocation method. PF-JUGPA first pre-selects users via a PF metric combining instantaneous rate proxies and historical averages, then performs fairness-aware grouping and power allocation by maximizing a weighted sum rate whose weights are inversely proportional to historical service rates. Simulation results show that PF-JUGPA significantly improves the Jain fairness index and weak-user average rates with only a modest sum-rate loss compared to sum-rate-oriented and round-robin baselines. The findings confirm that embedding long-term service history into both scheduling and resource allocation yields an effective throughput--fairness--sensing tradeoff in uplink NOMA-ISAC.
\end{abstract}


\section{Introduction}
Integrated sensing and communication (ISAC) has been widely recognized as a key technology for 6G networks, since it enables communication and sensing functions to share spectrum, hardware, and signal processing resources [1]--[3]. Meanwhile, NOMA improves spectral efficiency and connectivity by allowing multiple users to access the same time-frequency resource block [4]. The combination of ISAC and NOMA therefore provides a promising framework for future wireless systems. In particular, in uplink NOMA-ISAC, multiple users simultaneously transmit to a sensing-capable base station (BS), making user grouping and power allocation under sensing constraints a fundamental design problem [5]--[7].

Existing NOMA-ISAC resource allocation schemes mainly maximize the instantaneous communication sum rate while satisfying sensing requirements [6], [8]. Although effective in improving short-term throughput, such designs tend to repeatedly favor users with strong instantaneous channels. In practical systems with a large candidate user pool, i.e., $K \gg L$, where only $L$ users can be scheduled in each slot, this throughput-oriented strategy may lead to persistent long-term service imbalance. Users with weak channels or poor past service are likely to be scheduled less frequently and thus suffer from poor long-term throughput. This issue becomes even more severe when sensing constraints are tightened, since communication resources are further restricted.

Proportional fairness (PF) is a classical mechanism for balancing instantaneous transmission opportunities and historical service status, and has been widely used to improve long-term fairness in wireless networks [9]. However, directly applying conventional PF scheduling to uplink NOMA-ISAC is nontrivial. Unlike orthogonal access systems, uplink NOMA requires coupled decisions on user selection, user grouping, and power allocation, while the sensing function further imposes additional quality constraints. Hence, fairness enhancement should be incorporated not only into the scheduling layer but also into the per-slot resource allocation process. Although fairness-aware ISAC designs have been studied in non-NOMA settings [10], long-term fairness for uplink NOMA-ISAC remains insufficiently explored, especially when both scheduling and joint grouping/power allocation are considered under sensing constraints.

Motivated by the above observations, this letter proposes PF-JUGPA, a proportional-fair scheduling based joint user grouping and power allocation method for uplink NOMA-ISAC. The key idea is to embed historical service awareness into two coupled stages. First, a PF-based pre-scheduling metric is introduced to select users according to both instantaneous rate proxies and historical average service rates, thereby improving the transmission opportunities of historically underserved users. Second, for the scheduled user set, a fairness-aware joint grouping and power allocation stage is developed by maximizing a weighted sum rate, where the user weights are inversely proportional to historical average service rates. In this way, users with poorer long-term service records are prioritized not only during user selection but also during instantaneous resource allocation.

The main contributions of this letter are summarized as follows. First, a PF-based pre-scheduling mechanism is developed for uplink NOMA-ISAC to alleviate long-term service imbalance in large-user scheduling scenarios. Second, a fairness-aware joint user grouping and power allocation problem is formulated under sensing constraints, where historical average service rates are explicitly incorporated into the objective through dynamic user weights. Third, a low-complexity two-stage framework is proposed by combining PF user pre-selection, candidate grouping generation, and fairness-driven power allocation. Finally, simulation results show that the proposed PF-JUGPA scheme significantly improves the Jain fairness index and the average rate of weak users, while incurring only a modest loss in sum communication rate and maintaining sensing reliability.

\section{System Model and Performance Metrics}
\subsection{System Model}
Consider an uplink NOMA-ISAC system. The BS is equipped with $M$ receive antennas, and there are $K$ single-antenna communication users and $Q$ targets to be sensed in the system. In each time slot, the BS selects $L$ users from the entire user set to participate in uplink transmission and partitions them into $U$ NOMA groups. The BS performs target sensing while receiving uplink communication signals. To highlight the fairness scheduling and resource allocation mechanisms, we adopt an equivalent scalar channel model, where $h_k$ denotes the post-combining equivalent channel gain of user $k$. Here, $h_k$ absorbs array processing gains, beamforming/receive combining gains, and the corresponding large-scale and small-scale fading effects.

Let $\mathcal S(t)$ denote the set of users scheduled in time slot $t$, satisfying $|\mathcal S(t)|=L$. For any NOMA group $u$, let $\mathcal G_u(t)$ denote its set of users. Within each group, successive interference cancellation (SIC) is employed, and decoding is performed in descending order of received power. For user $k\in\mathcal G_u(t)$ in group $u$, the received signal-to-interference-plus-noise ratio (SINR) is defined as
\begin{equation}
\gamma_k(t)
=
\frac{p_k(t)h_k(t)}{\sum\limits_{i\in\mathcal I_k(t)}p_i(t)h_i(t)+I_u^{\mathrm{s}}(t)+\sigma^2},
\label{eq:sinr_user}
\end{equation}
where $p_k(t)$ is the transmit power of user $k$, $\mathcal I_k(t)$ denotes the set of co-group users that are ordered after user $k$ in the SIC sequence and have not yet been canceled, $I_u^{\mathrm{s}}(t)$ represents the equivalent interference term caused by sensing occupation or sensing coupling associated with group $u$, and $\sigma^2$ is the post-combining equivalent noise power. To avoid model ambiguity, we do not introduce the uncontrollable variable ``target transmit power''; instead, the impact of sensing on communication is uniformly abstracted into $I_u^{\mathrm{s}}(t)$, and the impact of communication on sensing is uniformly reflected in the subsequent sensing SINR.
Thus, the instantaneous achievable rate of user $k$ in time slot $t$ is
\begin{equation}
R_k(t)=\log_2\big(1+\gamma_k(t)\big).
\label{eq:rate_user}
\end{equation}

For target $q$, its equivalent sensing SINR is defined as
\begin{equation}
\gamma_q^{\mathrm r}(t)=\frac{\eta_q(t)}{\zeta_q(t)+\sigma_{\mathrm r}^2},
\label{eq:radar_sinr}
\end{equation}
where $\eta_q(t)$ denotes the effective signal power corresponding to the target echo, $\zeta_q(t)$ represents the equivalent interference power caused by communication multiplexing, echoes from other targets, and residual self-interference, and $\sigma_{\mathrm r}^2$ is the sensing-side noise power. To guarantee sensing quality, the system must satisfy the minimum sensing threshold constraint
\begin{equation}
\gamma_q^{\mathrm r}(t)\ge \phi_{\min},\quad \forall q=1,\ldots,Q.
\label{eq:sensing_constraint}
\end{equation}

\subsection{Long-Term Service Rate and Fairness Metrics}
To characterize cross-slot long-term fairness, we introduce the historical average service rate of a user:
\begin{equation}
\bar R_k(t+1)=(1-\beta)\bar R_k(t)+\beta x_k(t)R_k(t),
\label{eq:avg_rate_update}
\end{equation}
where $\beta\in(0,1)$ is the forgetting factor, and $x_k(t)\in\{0,1\}$ indicates whether user $k$ is scheduled in time slot $t$. When a user is not scheduled, $x_k(t)=0$, and its long-term average service level naturally decays. Equation \eqref{eq:avg_rate_update} reflects the average service status of a user over long-term operation and will play a role in both scheduling and resource allocation.

To quantify the system's long-term fairness, we adopt the Jain fairness index:
\begin{equation}
\mathcal J(t)=\frac{\left(\sum_{k=1}^{K}\bar R_k(t)\right)^2}{K\sum_{k=1}^{K}\bar R_k^2(t)}.
\label{eq:jain}
\end{equation}
$\mathcal J(t)\in(0,1]$, and a value closer to $1$ indicates a more balanced long-term average service among all users.

Furthermore, to more stably measure the performance of weak user groups, we define the average rate of the weakest $\rho$-proportion users. Let $\mathcal W_\rho$ denote a fixed weak user set consisting of the bottom $\rho K$ users sorted by their large-scale channel gains in ascending order. Then, their long-term average group rate is written as
\begin{equation}
\bar R_{\mathrm{tail}}(t)=\frac{1}{|\mathcal W_\rho|}\sum_{k\in\mathcal W_\rho}\bar R_k(t).
\label{eq:tail_rate}
\end{equation}
In the simulation section, we set $\rho=0.2$.

\section{PF-JUGPA: Proportional-Fairness Based Joint Resource Allocation}
\subsection{Proportional-Fair Pre-Scheduling}
In the user selection stage of time slot $t$, to balance instantaneous link quality and historical service status, we construct a PF metric for each candidate user:
\begin{equation}
\mu_k(t)=\frac{\tilde R_k(t)}{\bar R_k(t)+\epsilon},
\label{eq:pf_metric}
\end{equation}
where $\epsilon>0$ is a small constant to avoid division by zero, and $\tilde R_k(t)$ is an instantaneous achievable rate proxy for user $k$. To control the complexity of pre-scheduling, we adopt the single-user approximate rate:
\begin{equation}
\tilde R_k(t)=\log_2\!\left(1+\frac{p_{c,\max}h_k(t)}{\sigma^2}\right).
\label{eq:proxy_rate}
\end{equation}
This proxy is used only for coarse screening among all users; the impact of intra-NOMA-group interference and sensing constraints on the true rate will be explicitly corrected in the second-stage joint grouping and power allocation. The BS selects the top $L$ users with the largest $\mu_k(t)$ from all $K$ users to form the scheduling set $\mathcal S(t)$ for the current time slot.

The intuitive explanation of \eqref{eq:pf_metric} is that users with better instantaneous link quality tend to be selected due to larger $\tilde R_k(t)$, while users with lower historical average service rates also receive higher priority due to a smaller denominator. Thus, this mechanism can compensate for underserved users over the long term and mitigate the phenomenon of strong users persistently dominating.

\subsection{Fairness-Aware Joint User Grouping and Power Allocation}
Given the scheduling set $\mathcal S(t)$, we further explicitly inject long-term fairness information into the grouping and power control stage. The fairness weight of user $k$ is defined as
\begin{equation}
\omega_k(t)=\frac{1}{\bar R_k(t)+\epsilon}.
\label{eq:weight}
\end{equation}
This weight is directly determined by the user's historical average service rate: the less historical service, the larger the weight, thus gaining more attention in the current slot's resource allocation. Based on this, we solve the following problem in the resource allocation stage of time slot $t$:
\begin{subequations}
\begin{align}
\max_{\{\mathcal G_u(t)\},\,\{p_k(t)\}} \quad &
\sum_{k\in\mathcal S(t)}\omega_k(t)R_k(t)
\label{prob:main_obj}\\
\text{s.t.}\quad
& R_k(t)\ge R_{\min},\quad \forall k\in\mathcal S(t),
\label{cons:rmin}\\
& 0\le p_k(t)\le p_{c,\max},\quad \forall k\in\mathcal S(t),
\label{cons:power}\\
& \sum_{k\in\mathcal S(t)} p_k(t)\le P_{\mathrm{tot}},
\label{cons:ptot}\\
& \gamma_q^{\mathrm r}(t)\ge \phi_{\min},\quad \forall q=1,\ldots,Q,
\label{cons:sensing}\\
& \bigcup_{u=1}^{U}\mathcal G_u(t)=\mathcal S(t),\; \mathcal G_u(t)\cap\mathcal G_v(t)=\emptyset,\ u\neq v.
\label{cons:group}
\end{align}
\end{subequations}

The objective function \eqref{prob:main_obj} essentially maximizes the weighted sum rate of the current time slot, where the weights are induced by long-term average service rates. Compared with an unweighted sum-rate objective, this form is more unified and easier to interpret: it prioritizes the incremental benefit of users with poorer historical service in the current slot, thus remaining consistent with the long-term fairness philosophy of PF.

\subsection{Low-Complexity Two-Stage Solution Framework}
Since problem \eqref{prob:main_obj}--\eqref{cons:group} involves both combinatorial grouping variables and continuous power variables, directly obtaining the globally optimal solution is computationally expensive. Therefore, we adopt a low-complexity two-stage solution framework.

\textbf{1) Candidate Grouping Generation:}  
For the scheduled user set $\mathcal S(t)$, users are first sorted in descending order of their equivalent channel gains. Then, a strong-weak pairing principle is adopted to generate a set of candidate groupings, i.e., preferentially assigning strong users and weak users to the same NOMA group to balance SIC separability and fairness gains. In the considered setting with $L=4$ and $U=2$, each group contains two users, and thus only a limited number of candidate grouping schemes need to be compared.

\textbf{2) Power Allocation and Scheme Selection:}  
For each candidate grouping scheme, a fairness-weight-based power control rule is applied: subject to constraints \eqref{cons:rmin}--\eqref{cons:sensing}, relatively higher transmit power is allocated to users with lower historical average service rates. Subsequently, the weighted sum-rate objective is computed, and the optimal candidate grouping and its corresponding power configuration are selected as the final resource allocation result for the current time slot.

For ease of implementation, we adopt the following normalized power allocation form:
\begin{equation}
p_k(t)=\min\!\left\{p_{c,\max},\;\frac{\omega_k(t)}{\sum_{i\in\mathcal S(t)}\omega_i(t)}P_{\mathrm{tot}}\right\},
\label{eq:power_rule}
\end{equation}
where $P_{\mathrm{tot}}$ is the total communication power budget for the current time slot. If a candidate grouping cannot satisfy the sensing threshold constraint under this allocation, constraint correction is performed; if it remains infeasible after correction, the candidate grouping is discarded.

The algorithm flow is shown in Algorithm~1.

\begin{algorithm}[t]
\caption{PF-JUGPA}
\label{alg:pfjugpa}
\small
\begin{algorithmic}[1]
\STATE Initialize $\{\bar R_k(0)\}_{k=1}^K$
\WHILE{each time slot $t=1,2,\ldots$}
    \STATE Compute $\tilde R_k(t)$ according to \eqref{eq:proxy_rate}
    \STATE Compute PF metric $\mu_k(t)$ according to \eqref{eq:pf_metric}
    \STATE Select top $L$ users to form scheduling set $\mathcal S(t)$
    \STATE Generate candidate grouping set for $\mathcal S(t)$
    \FOR{each candidate grouping scheme}
        \STATE Compute fairness weights $\omega_k(t)$ according to \eqref{eq:weight}
        \STATE Allocate power according to \eqref{eq:power_rule} and check constraints
        \STATE Compute weighted sum-rate objective
        \STATE \hspace{1em} $\sum_{k\in\mathcal S(t)}\omega_k(t)R_k(t)$
    \ENDFOR
    \STATE Select the feasible grouping and power configuration
    \STATE \hspace{1em} with the maximum objective value
    \STATE Update $\bar R_k(t+1)$ according to \eqref{eq:avg_rate_update}
\ENDWHILE
\end{algorithmic}
\end{algorithm}

Structurally, PF+Fixed only retains the first-stage PF pre-scheduling and adopts fixed grouping and equal power allocation in the second stage, whereas PF-JUGPA further utilizes historical service status to perform fairness-aware grouping and power control in the second stage, thus better enabling the synergy between long-term fairness and instantaneous resource allocation.

\section{Simulation Results and Analysis}
In this section, Monte Carlo simulations are conducted to evaluate the performance of the proposed PF-JUGPA scheme. Unless otherwise specified, the system parameters are set as follows: the number of users is $K=32$, the number of scheduled users per slot is $L=4$, the number of sensing targets is $Q=6$, the number of NOMA groups is $U=2$, and the number of BS antennas is $M=5$. To ensure a fair comparison across different horizontal-axis parameter settings, the large-scale topology of users and targets is fixed within each simulation batch, while only the small-scale fading realizations are independently generated. Moreover, the forgetting factor is set to $\beta=0.05$, the minimum communication rate threshold is set to $R_{\min}=0$, and the weak-user proportion is set to $\rho=0.2$.

For performance comparison, the following benchmark schemes are considered. Here, the label ``OA'' in the figure legends denotes the corresponding per-slot resource allocation stage applied after user selection: 1) MaxSNR+OA: Users with the strongest instantaneous channel gains are scheduled, followed by sum-rate-oriented resource allocation over the selected users. 2) RR+OA: Users are scheduled in a round-robin manner, and then resource allocation is performed over the scheduled set. 3) PF+Fixed: PF-based pre-scheduling is adopted, while fixed user grouping and equal power allocation are used in the second stage. 4) PF-JUGPA: The proposed scheme, which combines PF-based pre-scheduling with fairness-aware joint user grouping and power allocation.

\begin{figure}[!t]
\centering
\includegraphics[width=2.9in]{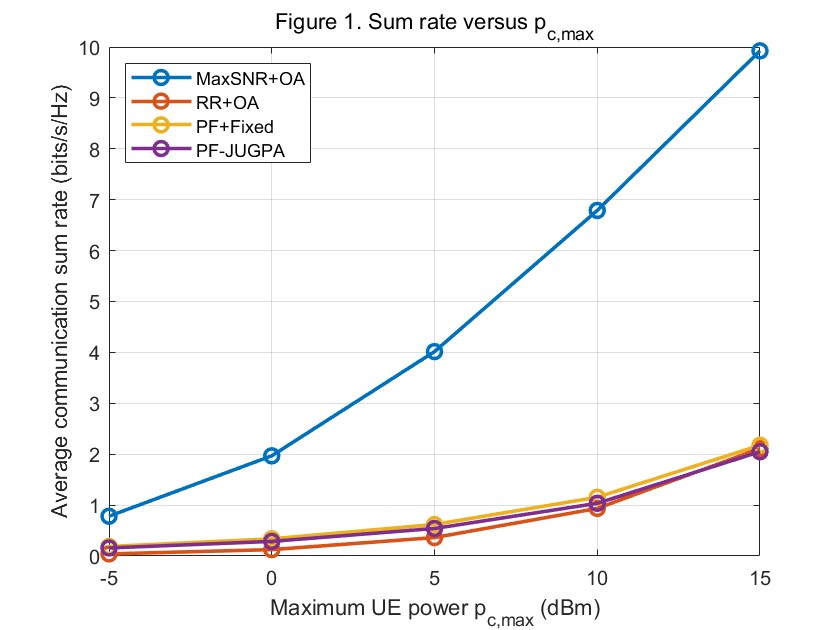}
\caption{Sum rate versus $p_{c,\max}$.}
\end{figure}

\subsection{Sum Communication Rate}
Fig.~1 shows the sum communication rate versus the maximum user transmit power $p_{c,\max}$. As expected, the sum communication rate of all schemes increases with $p_{c,\max}$, since a larger power budget directly improves the achievable uplink SINR. Among all schemes, MaxSNR+OA consistently achieves the highest sum rate over the entire power range, because it always prioritizes users with favorable instantaneous channels and performs throughput-oriented resource allocation. In contrast, RR+OA yields the lowest sum rate due to the lack of channel-aware scheduling gain. The two PF-based schemes lie between these two extremes. In particular, PF-JUGPA achieves a sum rate very close to that of PF+Fixed, with only a slight degradation. This result indicates that the proposed scheme does not seek to maximize the instantaneous sum rate alone; instead, it sacrifices only a limited amount of throughput in exchange for substantially improved long-term fairness and weak-user performance. Therefore, PF-JUGPA provides a more balanced throughput--fairness tradeoff than purely sum-rate-driven designs.

\subsection{Long-Term Fairness}
Fig.~2 illustrates the Jain fairness index versus $p_{c,\max}$. It can be observed that MaxSNR+OA consistently exhibits the lowest fairness, and its performance changes little as $p_{c,\max}$ increases. This confirms that selecting users mainly according to instantaneous channel strength leads to persistent long-term service imbalance. RR+OA improves the fairness level compared with MaxSNR+OA, since round-robin scheduling ensures that all users can obtain transmission opportunities over time. However, its fairness performance is still clearly inferior to that of the PF-based schemes, because it does not exploit historical service information in a utility-driven manner. Both PF+Fixed and PF-JUGPA achieve significantly higher Jain fairness indices, and the fairness performance further improves as $p_{c,\max}$ increases. More importantly, PF-JUGPA consistently outperforms PF+Fixed over the entire power range, showing that fairness enhancement at the scheduling stage alone is not sufficient. By further incorporating historical service information into the grouping and power allocation stage, PF-JUGPA more effectively balances the long-term service levels among users.

\begin{figure}[!t]
\centering
\includegraphics[width=3in]{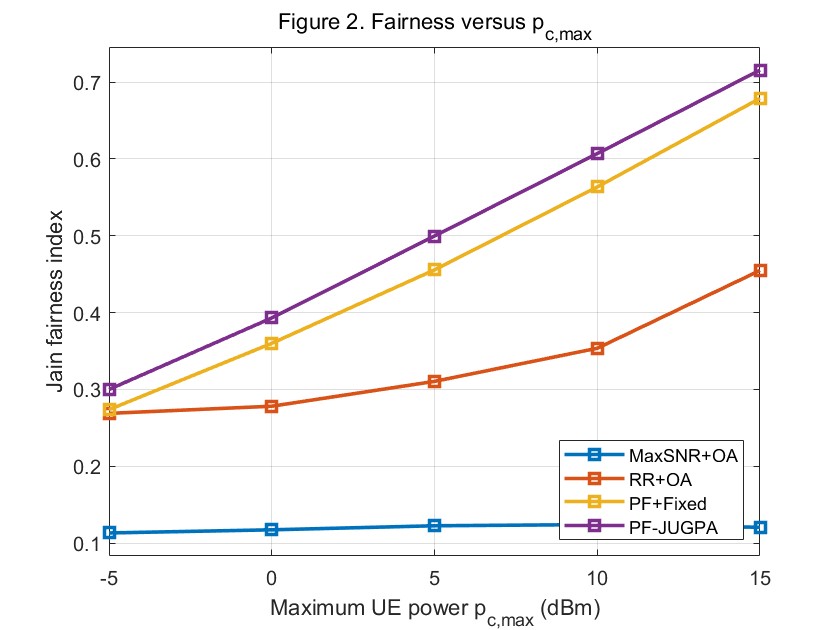}
\caption{Fairness versus $p_{c,\max}$.}
\end{figure}

\subsection{Weak User Group Performance}
Fig.~3 presents the average rate of the bottom 20\% users versus the sensing threshold $\phi_{\min}$. It is observed that the average weak-user rate of all schemes gradually decreases as $\phi_{\min}$ increases. This trend is expected, since stricter sensing requirements leave less flexibility for communication-oriented resource allocation. MaxSNR+OA performs the worst over the entire threshold range, which verifies that throughput-oriented designs tend to sacrifice weak users in order to favor users with better channels. RR+OA provides a moderate improvement over MaxSNR+OA, but it remains noticeably worse than the PF-based schemes. Both PF+Fixed and PF-JUGPA maintain substantially higher weak-user rates under most sensing thresholds, demonstrating the effectiveness of PF-based fairness enhancement. In addition, PF-JUGPA consistently and slightly outperforms PF+Fixed across the considered sensing-threshold range. This result shows that the proposed scheme not only improves the overall long-term fairness, but also more effectively protects the service quality of disadvantaged user groups under sensing-constrained operation.

\subsection{Sensing Performance}
Fig.~4 shows the detection probability $P_D$ versus the false alarm probability $P_{fa}$ for the proposed PF-JUGPA scheme. It can be observed that, as the sensing threshold $\phi_{\min}$ increases, the detection curve shifts upward, indicating improved sensing reliability under stricter sensing requirements. This behavior is consistent with the intended role of the sensing constraint, namely, enforcing more sensing-favorable resource allocation when higher sensing quality is required. Together with the communication and fairness results discussed above, Fig.~4 indicates that PF-JUGPA can maintain controllable sensing reliability under different sensing-threshold requirements while improving long-term fairness and weak-user performance.

Overall, the proposed PF-JUGPA scheme does not necessarily achieve the best performance in every single metric. However, it delivers a more favorable overall tradeoff among sum communication rate, long-term fairness, weak-user performance, and sensing reliability. This balanced behavior is the main advantage of PF-JUGPA over conventional sum-rate-oriented resource allocation strategies for uplink NOMA-ISAC systems.

\begin{figure}[!t]
\centering
\includegraphics[width=2.9in]{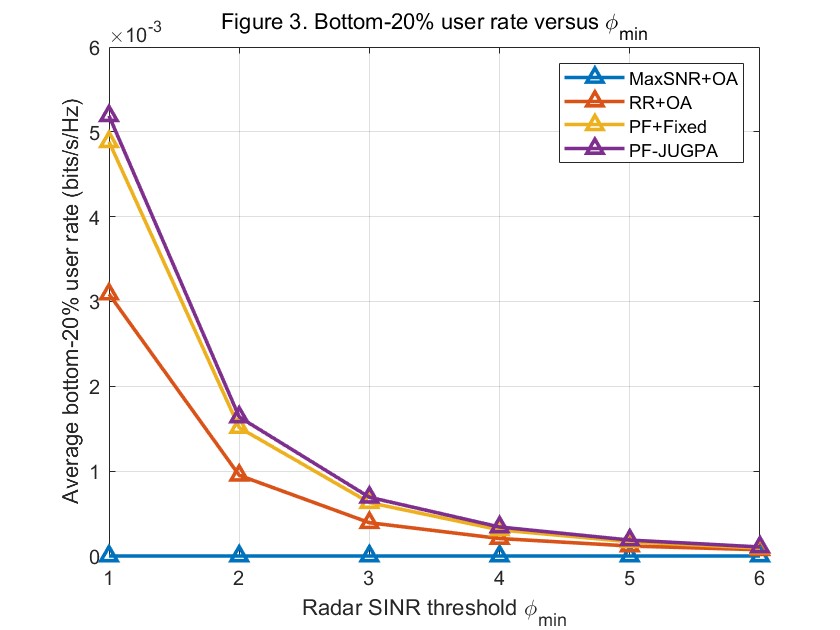}
\caption{Bottom-20\% user rate versus $\phi_{\min}$.}
\end{figure}

\section{Conclusion}
This letter has investigated the issue of insufficient long-term fairness in uplink NOMA-ISAC systems and proposed a two-stage resource allocation framework, PF-JUGPA, based on proportional-fair pre-scheduling and fairness-aware joint user grouping and power allocation. Unlike existing designs that primarily focus on maximizing instantaneous sum rate, this work incorporates historical service status information into both the temporal scheduling layer and the per-slot resource allocation layer. Specifically, the PF-based pre-scheduling mechanism alleviates long-term service imbalance by improving the transmission opportunities of historically underserved users, while the fairness-aware grouping and power allocation stage further enhances their instantaneous resource gains through dynamic weights determined by historical average service rates.

Simulation results show that the proposed method can significantly improve the Jain fairness index and the average rate of weak user groups, while incurring only a limited loss in the overall sum communication rate. In addition, the sensing results indicate that PF-JUGPA can maintain sensing reliability under different sensing-threshold requirements, demonstrating that fairness enhancement is achieved without losing the basic sensing capability of the ISAC system. These results confirm that introducing long-term service awareness into both scheduling and resource allocation is an effective way to achieve a more balanced tradeoff among communication efficiency, long-term fairness, weak-user protection, and sensing performance in uplink NOMA-ISAC networks.

Overall, PF-JUGPA provides a low-complexity and practically meaningful design framework for fairness-oriented uplink NOMA-ISAC resource management. Future work may consider more rigorous sensing signal models, robust designs under imperfect channel state information and sensing uncertainty, as well as learning-assisted online low-complexity algorithms for dynamic large-scale systems.

\begin{figure}[!t]
\centering
\includegraphics[width=3in]{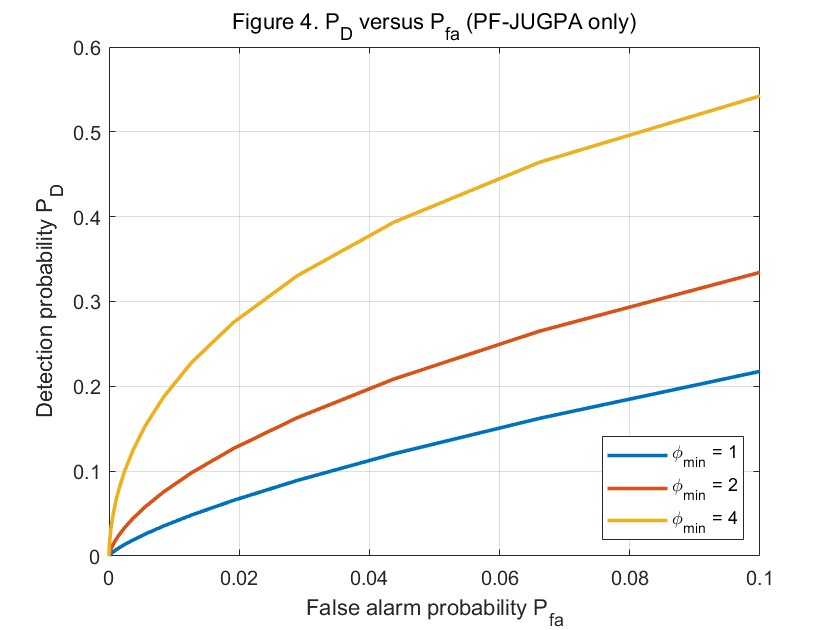}
\caption{$P_D$ versus $P_{fa}$ (PF-JUGPA only).}
\end{figure}


\end{document}